\pdfoutput=1
\documentclass[11pt,twoside,a4paper,cmspaper,final,collab]{cms-tdr}
\def\svnVersion{298000b}\def\svnDate{2023/09/05}\def\cmsCernNoTag{CERN-EP-2021-071}\def\cmsCernDate{\today}\def\cmsMessage{Published in Physical Review Letters as \href{http://dx.doi.org/10.1103/PhysRevLett.131.051901}{\doi{10.1103/PhysRevLett.131.051901}.}}
\begin{document}\cmsNoteHeader{HIN-18-011}

\newcommand{\QT}{Q_\text{T}}
\newcommand{\myPT}{P_\text{T}}
\newcommand{\vQT}{\vec{Q}_\text{T}}
\newcommand{\vPT}{\vec{P}_\text{T}}
\newcommand{\vpta}{\vec{p}_\text{T,1}}
\newcommand{\vptb}{\vec{p}_\text{T,2}}
\newcommand{\dnf}{\Delta\eta^{F}}
\newcommand{\dnb}{\Delta\eta^{B}}
\newcommand{\etam}{\eta_{\text{max}}}
\newcommand{\Avcos}{\langle \cos(2\Phi) \rangle}
\newcommand{\dndphi}{\rd N/\rd\Phi}
\newcommand{\RAPGAP}{\textsc{rapgap}\xspace}

\newlength\cmsFigWidth
\ifthenelse{\boolean{cms@external}}{\setlength\cmsFigWidth{0.49\textwidth}}{\setlength\cmsFigWidth{0.65\textwidth}}
\ifthenelse{\boolean{cms@external}}{\providecommand{\cmsLeft}{upper\xspace}}{\providecommand{\cmsLeft}{left\xspace}}
\ifthenelse{\boolean{cms@external}}{\providecommand{\cmsRight}{lower\xspace}}{\providecommand{\cmsRight}{right\xspace}}
\cmsNoteHeader{HIN-18-011} 
\title{Azimuthal Correlations within Exclusive Dijets with Large Momentum Transfer in Photon-Lead Collisions}

\date{\today}

\abstract{
The structure of nucleons is multidimensional and depends on the transverse momenta, spatial geometry, and polarization of the constituent partons.  Such a structure can be studied using high-energy photons produced in ultraperipheral heavy-ion collisions. The first measurement of the azimuthal angular correlations of exclusively produced events with two jets in photon-lead interactions at large momentum transfer is presented, a process that is considered to be sensitive to the underlying nuclear gluon polarization. This study uses a data sample of ultraperipheral lead-lead collisions at $\sqrtsNN = 5.02\TeV$, corresponding to an integrated luminosity of 0.38\nbinv, collected with the CMS experiment at the LHC. The measured second harmonic of the correlation between the sum and difference of the two jet transverse momentum vectors is found to be positive, and rising, as the dijet transverse momentum increases. A well-tuned model that has been successful at describing a wide range of proton scattering data from the HERA experiments fails to describe the observed correlations, suggesting the presence of gluon polarization effects.

}

\hypersetup{%
pdfauthor={CMS Collaboration},%
pdftitle={Azimuthal correlations within exclusive dijets with large momentum transfer in photon-lead collisions},%
pdfsubject={CMS},%
pdfkeywords={CMS, upc, dijets, hard exclusive}}
\maketitle

Considerable experimental and theoretical effort is being devoted to the study of the momentum distribution of gluons in 
nuclei~\cite{Albacete:2014fwa,AbdulKhalek:2021gbh}. The aim of such studies is to determine whether the distribution of 
low momentum gluons saturates, constituting a new state of matter. Accumulating evidence since early studies at BNL RHIC~\cite{BRAHMS:2004xry,STAR:2006dgg,PHENIX:2011puq,PHOBOS:2009sau} 
points to a reduction of the density of low momentum gluons in heavy nuclei, as compared to the density in individual protons, 
at high energies. While experimental work has so far focused on one-dimensional studies, the gluon distribution  is intrinsically multidimensional, 
depending on radial distance to the nucleus center, transverse momentum and polarization~\cite{Diehl:2015uka}.

Recent theoretical studies relate the polarization of gluons within nuclei to angular correlations of final-state jets produced in photon-hadron interactions ~\cite{Altinoluk:2015dpi, Hatta:2016dxp, Hagiwara:2017fye, Hatta:2017cte, Mantysaari:2019csc, Dumitru:2018kuw, Hagiwara:2016kam, Hagiwara:2017ofm, Zhou:2016rnt,Strikman:2005yv, Guzey:2016tek, Kotko:2017oxg}. The cross section for diffractive dijet production depends on the orientation of the quark-antiquark color dipole with respect to the radial distance. This orientation dependence results in an azimuthal decorrelation of the nominally back-to-back jets that comprise the dijet. The nature of gluon polarization in nuclei is of great interest, yet remains almost completely unexplored. While the gluon polarization density matrices have been theoretically classified, estimates of the magnitude of gluon polarization effects vary widely~\cite{Altinoluk:2015dpi, Hatta:2016dxp, Hagiwara:2017fye, Hatta:2017cte, Mantysaari:2019csc, Dumitru:2018kuw,Hagiwara:2016kam, Hagiwara:2017ofm, Zhou:2016rnt}. Theory predicts that the distribution of elliptically polarized gluons~\cite{Mantysaari:2019csc} can be probed with the second Fourier moment $\Avcos$ of the distribution of the azimuthal angle $\Phi$ between the sum and the difference of the two jet transverse momentum vectors of a dijet~\cite{Hatta:2016dxp}. In this approach, the observed moment measures a fundamental correlation of gluon polarizations with their transverse momenta. 

After a preliminary version of the current results was shown, a calculation based on final-state radiation was found to 
result in a nonzero value for this harmonic, giving an alternative explanation for any observed correlations~\cite{Hatta:2020bgy}. Initial-state radiation from parton showers could also lead to a decorrelation. Experimental measurements are essential to resolve the possible effects and to gain insight into the magnitude of gluon polarization within nuclei. 

Photon-nucleus interactions can be produced using ultrarelativistic heavy-ion beams. At the CERN Large Hadron Collider, 
high-energy lead (Pb) beams produce a large flux of virtual photons that can interact with an 
oncoming lead nucleus~\cite{Baltz:2007kq, Contreras:2015dqa,Klein:2020fmr}, in so-called ultraperipheral heavy-ion collisions (UPCs). The CMS Collaboration has recently studied UPC exclusive vector meson photoproduction and photon-photon 
scattering~\cite{Khachatryan:2016qhq,Sirunyan:2019nog, Sirunyan:2018sav,Sirunyan:2018fhl}. The ALICE, ATLAS, and LHCb Collaborations have also recently reported on UPC measurements~\cite{Abelev:2012ba, Aad:2019ock, LHCb:2018rcm}. The CDF and ZEUS Collaborations have studied exclusive dijet production in proton-antiproton~\cite{Aaltonen:2007hs} and diffractive electron-proton collisions~\cite{Abramowicz:2015vnu}, respectively. While there are no published results on dijet production in photon-nucleus interactions, this process has generated considerable interest in the context of the future Electron-Ion Collider~\cite{AbdulKhalek:2021gbh}.

This Letter reports on the first measurement of $\Avcos$ for exclusive dijet events produced at large momentum transfer and with a large rapidity gap in photon-lead collisions. In such events, the dijet and the incoming photon are typically moving in the same direction, and are produced when the incoming photon fluctuates into a small color dipole, in comparison to the size of the nucleus. This dipole can then probe the interior of the Pb nucleus. Tabulated results are provided in the HEPData record for this analysis~\cite{hepdata}.

The central feature of the CMS apparatus is a superconducting solenoid of 6\unit{m} internal diameter, providing a magnetic field of 3.8\unit{T}. 
Within the solenoid volume are a silicon pixel and strip tracker, a lead tungstate crystal electromagnetic calorimeter (ECAL), 
and a brass and scintillator hadron calorimeter (HCAL). ECAL provides coverage in pseudorapidity $\abs{\eta} < 1.48$ in a barrel region 
and $1.48 < \abs{\eta} < 3.0$ in two end cap (EE) regions, while HCAL covers  $\abs{\eta} < 1.3$ for the barrel and $1.3 <\abs{\eta} < 3.0$ 
for the two end cap (HE) sections, respectively. Detector elements in ECAL and HCAL are grouped into ``towers.'' 
Hadron forward (HF) calorimeters extend the $\eta$ coverage provided by the barrel and end cap detectors to $3.0 < \abs{\eta} < 5.2$. 
The data sample is collected with a two-level trigger system~\cite{Khachatryan:2016bia}: at the hardware level events are selected by 
custom hardware processors, while a subsequent software trigger uses fast versions of the off-line reconstruction code. A detailed description of the CMS detector, 
together with a definition of the coordinate system used and the relevant kinematic variables, can be found in Ref.~\cite{Chatrchyan:2008aa}.

The analysis uses a data sample of lead-lead 
collisions at $\sqrtsNN = 5.02\TeV$ corresponding to an integrated luminosity of 0.38\nbinv collected in fall 2015. 
The average number of inelastic PbPb collisions per bunch crossing is about 0.01. 
Events are selected with a dedicated trigger designed to record a wide variety of photon-lead processes. 
The hardware trigger requires a transverse energy of at least 5\GeV in any of the ECAL towers, and at least one of the two HF calorimeters 
is required to not have a signal above the noise threshold. The software-based trigger requires at least one reconstructed track in the pixel detector. 
Since the charged particle multiplicities are low, events are reconstructed in the same way as if they were \textit{pp} collisions~\cite{Sirunyan:2018qel}. 
The primary vertex is required to be within 20\cm of the nominal center of the detector along the beam direction.

By combining information from all subdetectors, a particle-flow (PF) algorithm~\cite{Sirunyan:2017ulk} attempts to identify 
all stable particles in an event, classifying them as electrons, muons, photons, and charged or neutral hadrons. 
Jets are reconstructed from these PF candidates using the anti-$\textit{k}_\textit{T}$ algorithm~\cite{Cacciari:2008gp} with a distance parameter of 0.4, 
as implemented in the FASTJET package~\cite{Cacciari:2011ma}. The jet energy is calibrated by applying a multiplicative factor to relate, 
on average, the energy of the detector jet to that of the corresponding generator-level particle jet (energy scale correction). 
The leading (subleading) jet is required to have a $\textit{p}_\textit{T} > 30(20)\GeV$.   

The symmetry of the PbPb system leads to dijet events where the jets can travel in either direction through CMS. 
For the events of interest, the dijet is expected to be predominantly in the photon direction. This is taken as the ``forward" direction. Events that have any additional jets with $\textit{p}_\textit{T} > 20\GeV$ in either the forward or backward direction are rejected. While the expected yield for three jet events is negligible, dijet events with a nonjet-related PF candidate of more than 6\GeV in the forward hemisphere are rejected. Events are discarded if they have an energy deposit in any of the calorimeter towers in the forward 
and backward rapidity hemispheres above the noise threshold (3.9/3.0/3.2\GeV in HF/HE/EE)~\cite{Sirunyan:2018fhl}. 

Diffractive dijets are characterized by events with a large rapidity gap. A pseudorapidity gap is defined with $\dnf =2.4-\etam$, where 2.4 is the upper limit of the tracker and $\etam$ 
is the maximum pseudorapidity of any high-purity charged track as defined in Ref.~\cite{trkpur} and having $\textit{p}_\textit{T} > 0.2\GeV$. 
A similar backward pseudorapidity gap $\dnb$ is defined for the opposite $\eta$ hemisphere. 
It is required that $\dnb > \dnf$ and that  $\dnb$ is greater than 1.2, effectively selecting dijet events 
that are located in the forward direction with a large rapidity gap as measured using the tracker. Finally, one of the jets must be within one unit of pseudorapidity of the track defining the backward rapidity gap. Event samples with the dijet detected in either hemisphere of CMS are statistically independent and consistent 
with each other within 1\% for the $\Avcos$ value. The samples are combined using the convention for defining the forward direction as described above. This means that we invert the rapidity sign for the sample having a backward going dijet. A total of 6785 dijet events remain after applying the above requirements. 

In this analysis, the \RAPGAP (version 3.303)~\cite{Jung:1993gf} Monte Carlo (MC) generator, which was developed to explore
electron-proton (\textit{ep}) collisions, is used for comparisons with the experimental
results, for detector acceptance and resolution corrections, and for studying sources of systematic uncertainties. The exclusive dijet production model used in \RAPGAP is based on the boson-gluon fusion process for diffractive electron-proton collisions, supplemented with initial and final-state parton showers. The diffractive parton distribution functions are set to the ``H1 2006 fit-A" model. Finally, the generated photon  energy  spectrum has been reweighted to follow that of the Weizsacker-Williams photon flux distribution in lead nuclei described in Ref. ~\cite{Klein:2016yzr}. The simulated events are processed and reconstructed in the same way as the collision data. \GEANTfour~\cite{Agostinelli:2002hh} is used to account for the detector response. For the simulated events, 94\% have the interacting photon and the dijet in the same hemisphere and, of these, 99\% pass the selection criteria described above.

\begin{figure*}[ht!]
\centering
\includegraphics[width=\textwidth]{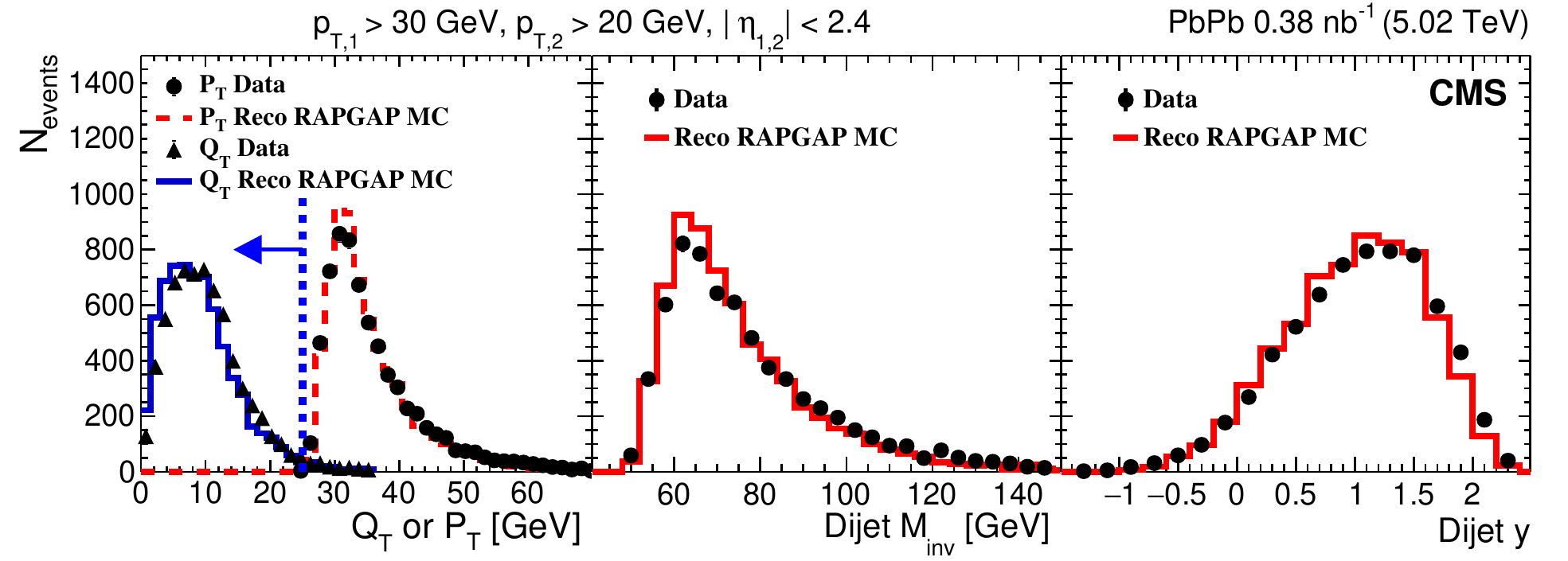}
\caption{Magnitudes of the vector sum ($\QT$) and vector difference ($\myPT$) of the two jets (left). 
	The dashed blue line illustrates the $\QT < 25 \GeV$ requirement. 
	Invariant mass (center) and rapidity (right) of the dijet candidates after all selection requirements. 
	The lines show the \RAPGAP MC generated events including  detector resolution effects. The statistical uncertainties
	are covered by the symbol size.\label{fig:qt_pt}}
\end{figure*}

Based on the transverse momenta of the two jets relative to the beam axis, $\vpta$ and $ \vptb$, the vector sum $\vQT$ and
vector difference $\vPT$ are defined, with $\vQT = \vpta + \vptb$ 
and $\vPT = \frac{1}{2} ( \vpta - \vptb)$~\cite{Altinoluk:2015dpi}. Theoretical calculations have focused on the ``back-to-back" regime, \ie $\myPT  >\QT$~\cite{Klein:2019qfb}. This requirement results 
in the rejection of 47 events (0.7\% of the sample). Figure~\ref{fig:qt_pt} shows the distributions of dijet $\QT$, $\myPT$, invariant mass, 
and rapidity for both data and reconstructed MC events. 
The good agreement between data and MC for the invariant mass and rapidity distributions suggests that the rescaling of 
the photon flux from \textit{ep} to PbPb is reasonable. The measured $\QT$ distribution has its maximum at 9\GeV, which is well within the large momentum transfer regime.   

To calculate the second moment of the azimuthal angle anisotropy, $\Avcos$, 
the angle $\Phi$ is obtained from the relation 
$\vPT\cdot\vQT = \abs{\vPT}\abs{\vQT} \cos(\Phi)$. 
For the angular correlation measurements, the $\vpta$ and $\vptb$ vectors used to calculate $\vQT$ and $\vPT$ 
are assigned randomly to the leading and subleading jets on an event-by-event basis. 
The dijet angular correlations are distorted compared to the underlying distributions by the acceptance and finite resolution of the detector. 
To correct for these effects, an unfolding procedure is used to estimate the  parton-level distributions. 
The TUnfold~\cite{Schmitt:2012kp} software package with the Tikhonov regularization method is used, 
where the strength of the regularization parameter is determined with the \textit{L}-curve scan method by minimizing the average global correlation coefficient. 
The response matrix is evaluated by using simulated dijet events that pass all of the analysis requirements. 
The unfolding of the $\dndphi$ distribution is performed 
for five equal intervals in $0 < \QT < 25\GeV$. The $\Avcos$ value is then calculated as a function of $\QT$. 

The following sources of systematic uncertainties on $\Avcos$ are considered: 
jet energy scale (JES) correction, JES nonclosure (JESNC), jet energy resolution (JER), jet $\eta$ resolution (JPR), 
jet azimuthal angular resolution (JAR), trigger efficiency (TR), and the purity estimation (PUR).
The uncertainty in the JES is 2\% for jets with $\textit{p}_\textit{T}$ of 20\GeV, as reported in Refs.~\cite{CMS:2015kaa, Sirunyan:2018qel}.
After the nominal correction factors are applied, a small nonclosure remains between the generator-level JES and that of the reconstructed jets. 
The systematic uncertainty from this effect is estimated by applying first a residual correction to the JES in simulation 
and then the dedicated selection requirements. 
The deviation from the nominal result is taken as an estimate of the systematic uncertainty. 
The simulation provides a JER of 14\% for 20\GeV jets, decreasing slightly as a function of jet $\textit{p}_\textit{T}$, 
consistent with Refs.~\cite{CMS:2015kaa, Sirunyan:2018qel}.
The JAR and JPR are found to be $0.025\pm0.005\stat$ and $0.015\pm0.005\stat$, respectively, and are almost independent 
of $\textit{p}_\textit{T}$ and $\eta$ in the phase space relevant to this measurement. 
The uncertainties related to JER and JAR are estimated from the differences between the  resolution in data and MC as done in Ref.~\cite{CMS:2015kaa}. 
For this estimate, the detector-level jets in the simulation are broadened in order to account for the observed resolution in data. 
The $\Avcos$ uncertainty related to the JPR is found to be negligible. 

The uncertainty related to the purity of the signal (PUR) is estimated by varying the nominal pseudorapidity gap requirement between 0 and 2, 
and by varying the requirement for the minimal distance in pseudorapidity between the track defining the backward pseudorapidity gap and the nearest jet, which has
a nominal value of 1.0, from 0.8 to 1.2. 
Increasing the threshold on the total energy of the PF candidates outside the jet cones from 6.0 to 6.6\GeV 
has a negligible effect on the purity uncertainty. 
The uncertainty associated with the trigger efficiency is related to the component that requires the transverse energy 
in one of the ECAL towers to be larger than 5\GeV. This requirement tends to slightly increase the fraction of electromagnetic 
energy in the jets. To account for this effect, simulated events are weighted by the trigger efficiency as a function of 
jet $\textit{p}_\textit{T}$ on an event-by-event basis. 
This weighting reproduces the effect of the trigger efficiency on the measured  $\Avcos$ values. 
The other elements forming the trigger, discussed above, are fully efficient. 
The uncertainties related to the calorimeter exclusivity 
selections and to the unfolding procedure are negligible. The various components of the $\Avcos$ distribution systematic uncertainties are summarized in Table~\ref{tab:sys_raw}. 

\begin{table}[htb!]
\centering
\topcaption{Table of $\Avcos$ systematic uncertainties (absolute values). The individual components are discussed in the text.}
\begin{scotch}{c  c  c  c  c  c  c  c}
$\QT$ $[\GeVns]$ & JES  & JESNC  & JER  & JAR  & PUR & TR  & Total \\
\hline
0–5 & 0.018&0.004 & 0.002 & 0.004 &  0.002 & 0.004 &0.019\\
5–10 & 0.012 &0.006& 0.005 & 0.003 & 0.002 & 0.003 &0.015\\
10–15 & 0.010 &0.008& 0.007 & 0.002 & 0.002 & 0.001& 0.014\\
15–20 & 0.009 & 0.008&0.014 & 0.002 &  0.002 & 0.001&0.018\\
20–25 & 0.005 & 0.018&0.056& 0.001 & 0.002 & 0.002&0.059\\
\end{scotch}
\label{tab:sys_raw}
\end{table}

\begin{figure}[htb!]
\centering
\includegraphics[width=0.49\textwidth]{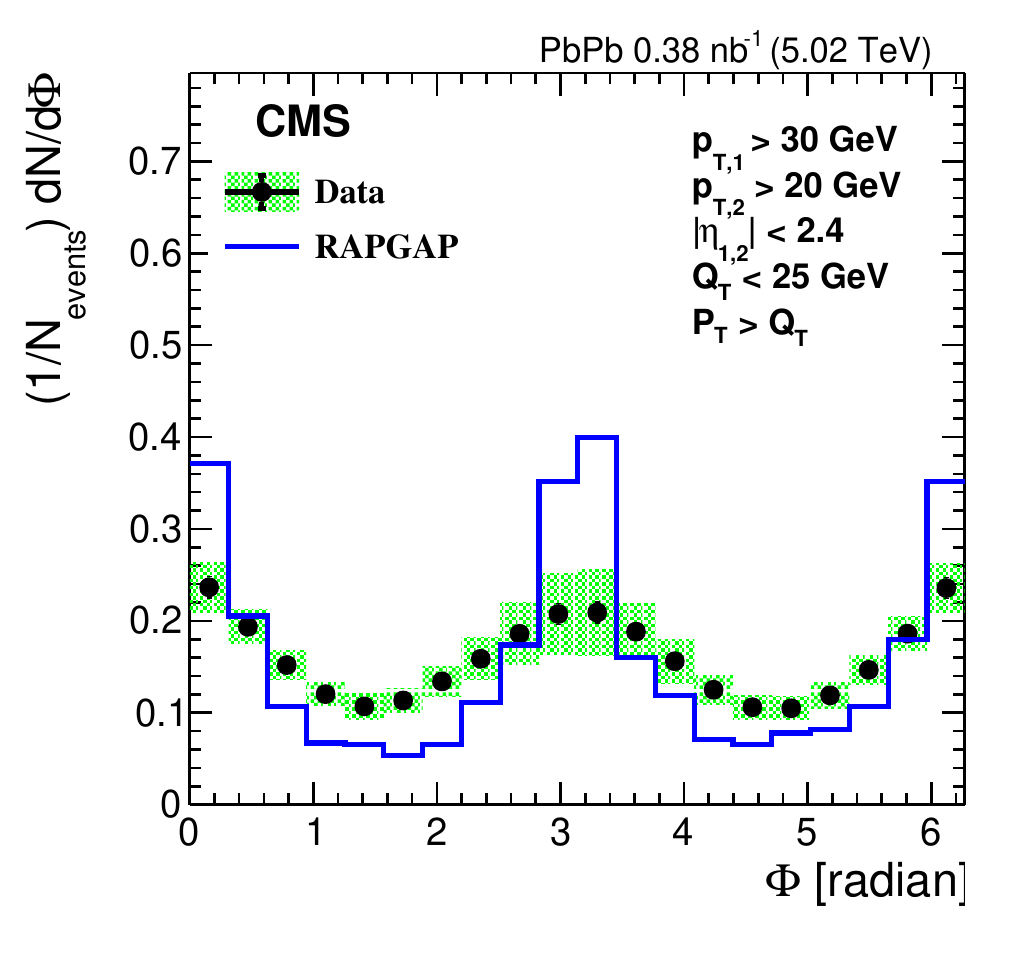}
\includegraphics[width=0.49\textwidth]{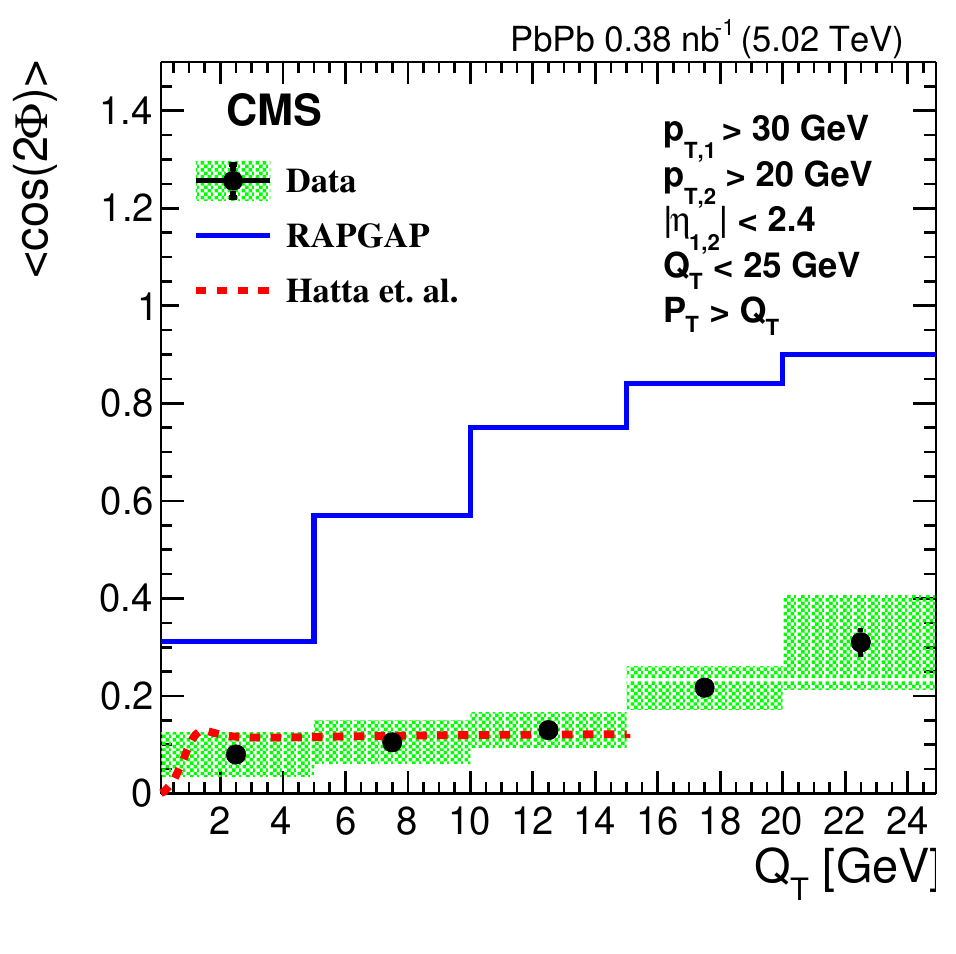}
\caption{The unfolded $1/{N_{\text{events}}} \,\dndphi$ distribution (\cmsLeft) and the unfolded $\Avcos$ values as a function of $\QT$ (\cmsRight). The corresponding distributions from the  \RAPGAP simulation at the generator level (blue lines) and theoretical calculation by Hatta \textit{et al.} \etal~\cite{Hatta:2020bgy} for $\Avcos$ (red dashed lines) are also shown for $\QT < 15 \GeV$, consistent with the back-to-back limit in their calculation. The dijet events are found predominantly in the forward direction, with ${0<\eta_\text{jet} < 2.4}$. Both the statistical (error bars) and systematic (green boxes) uncertainties are shown.\label{fig:v2_syst}}
\end{figure}

Figure~\ref{fig:v2_syst} (left) shows the unfolded angular distribution
$\dndphi$ for $\QT < 25\GeV$. The $\dndphi$ distribution peaks at 0 and $\pi$, implying a positive value of $\Avcos$. 
This phenomenon is not a trivial acceptance effect as demonstrated by performing the analysis upon pairs of jets from different events, yielding the expected negative value of $\Avcos$~\cite{Dumitru:2018kuw}. 
Figure 2 (right) shows that $\Avcos$ rises steadily with $\QT$. 
The data have been compared to two different calculations that ignore the effect of elliptically polarized gluons. A calculation by Hatta \textit{et al}.~\cite{Hatta:2020bgy}, which assumes soft gluon radiation from final-state jets as the dominant effect of the azimuthal anisotropy, has $\Avcos$ initially rising, but then plateauing for $\QT > 1 \GeV$. A calculation based on the \RAPGAP model~\cite{Jung:1993gf}, which is tuned to HERA results, predicts that $\Avcos$ rises with $\QT$, but overshoots the data by a factor of 3-5. The Dokshitzer--Gribov--Lipatov--Altarelli--Parisi based initial-state parton shower in \RAPGAP produces a finite transverse momentum of the incoming parton, and this leads to a decorrelation of the resulting dijet constituents. Such decorrelations are also observed in data for the azimuthal angular difference between two jets in proton-proton collisions~\cite{Kotko:2017oxg}, and are qualitatively described by parton shower production. The decorrelation observed here can also be driven by parton showers, but since heavy ions are involved, there could be further decorrelation effects at small Bjorken $x$, which are not simulated in \RAPGAP. Such broadening of the azimuthal distribution for dijet events has been linked to gluon saturation effects when comparing proton-lead to proton-proton events~\cite{vanHameren:2019ysa}. The broadening claimed in this analysis corresponds to the comparison between photon-lead (data) and photon-proton (\RAPGAP) interactions.

In summary, the exclusive production of two jets in photon-lead interactions with a large rapidity gap has been studied for the first time. The dijet events are characterized by a large momentum transfer. The second harmonic of the angular correlation between the sum and difference of the two jet transverse momenta is found to be positive and rising, with dijet transverse momentum in the measured range 0–25\GeV. An \textit{a posteriori} calculation that includes the effect of soft gluons from final-state radiation describes the average magnitude of the correlations, but does not rise with the magnitude of the dijet total momentum $\QT$ to the extent found with the CMS data. A model~\cite{Jung:1993gf} that has been successful at describing a wide range of proton scattering data from the HERA experiments overestimates the strength of the correlations, suggesting the presence of gluon polarization effects. This experiment calls for new theoretical calculations to quantify the theoretical implications and opens a new direction for probing the high-energy limit of strong interactions.

\begin{acknowledgments}
We congratulate our colleagues in the CERN accelerator departments for the excellent performance of the LHC and thank the technical and administrative staffs at CERN and at other CMS institutes for their contributions to the success of the CMS effort. In addition, we gratefully acknowledge the computing centers and personnel of the Worldwide LHC Computing Grid and other centers for delivering so effectively the computing infrastructure essential to our analyses. Finally, we acknowledge the enduring support for the construction and operation of the LHC, the CMS detector, and the supporting computing infrastructure provided by the following funding agencies: BMBWF and FWF (Austria); FNRS and FWO (Belgium); CNPq, CAPES, FAPERJ, FAPERGS, and FAPESP (Brazil); MES and BNSF (Bulgaria); CERN; CAS, MoST, and NSFC (China); MINCIENCIAS (Colombia); MSES and CSF (Croatia); RIF (Cyprus); SENESCYT (Ecuador); MoER, ERC PUT and ERDF (Estonia); Academy of Finland, MEC, and HIP (Finland); CEA and CNRS/IN2P3 (France); BMBF, DFG, and HGF (Germany); GSRI (Greece); NKFIH (Hungary); DAE and DST (India); IPM (Iran); SFI (Ireland); INFN (Italy); MSIP and NRF (Republic of Korea); MES (Latvia); LAS (Lithuania); MOE and UM (Malaysia); BUAP, CINVESTAV, CONACYT, LNS, SEP, and UASLP-FAI (Mexico); MOS (Montenegro); MBIE (New Zealand); PAEC (Pakistan); MES and NSC (Poland); FCT (Portugal); MESTD (Serbia); MCIN/AEI and PCTI (Spain); MOSTR (Sri Lanka); Swiss Funding Agencies (Switzerland); MST (Taipei); MHESI and NSTDA (Thailand); TUBITAK and TENMAK (Turkey); NASU (Ukraine); STFC (United Kingdom); DOE and NSF (U.S.).
\end{acknowledgments}

\bibliography{auto_generated}

\cleardoublepage \appendix\section{The CMS Collaboration \label{app:collab}}\begin{sloppypar}\hyphenpenalty=5000\widowpenalty=500\clubpenalty=5000
\cmsinstitute{Yerevan Physics Institute, Yerevan, Armenia}
{\tolerance=6000
A.~Tumasyan\cmsorcid{0009-0000-0684-6742}
\par}
\cmsinstitute{Institut f\"{u}r Hochenergiephysik, Vienna, Austria}
{\tolerance=6000
W.~Adam\cmsorcid{0000-0001-9099-4341}, T.~Bergauer\cmsorcid{0000-0002-5786-0293}, M.~Dragicevic\cmsorcid{0000-0003-1967-6783}, A.~Escalante~Del~Valle\cmsorcid{0000-0002-9702-6359}, R.~Fr\"{u}hwirth\cmsAuthorMark{1}\cmsorcid{0000-0002-0054-3369}, M.~Jeitler\cmsAuthorMark{1}\cmsorcid{0000-0002-5141-9560}, N.~Krammer\cmsorcid{0000-0002-0548-0985}, L.~Lechner\cmsorcid{0000-0002-3065-1141}, D.~Liko\cmsorcid{0000-0002-3380-473X}, I.~Mikulec\cmsorcid{0000-0003-0385-2746}, F.M.~Pitters, N.~Rad, J.~Schieck\cmsAuthorMark{1}\cmsorcid{0000-0002-1058-8093}, R.~Sch\"{o}fbeck\cmsorcid{0000-0002-2332-8784}, M.~Spanring\cmsorcid{0000-0001-6328-7887}, S.~Templ\cmsorcid{0000-0003-3137-5692}, W.~Waltenberger\cmsorcid{0000-0002-6215-7228}, C.-E.~Wulz\cmsAuthorMark{1}\cmsorcid{0000-0001-9226-5812}, M.~Zarucki\cmsorcid{0000-0003-1510-5772}
% ...
\par}
\vskip\cmsinstskip
\dag:~Deceased\\
$^{1}$Also at TU Wien, Vienna, Austria\\
% ...
\end{sloppypar}
\end{document}